# Temperature-Induced magneto-structural transformations in nickel crystals

Khisa Sh. Borlakov[*] and Albert Kh. Borlakov

*North Caucasian State Humanitarian and Technological Academy,*

*36 Stavropolskaya str., Cherkessk, Russia, 369001*

The magnetic and structural transformations in the nickel crystal within the temperature range 470 < T < 631 K is studied on the basis of general phenomenological theory of spin-orbit phase transitions. A new phase transition at $T_{ls}$ ≈570 K conditioned by cooperative relativistic interactions effect is prognosticated and the experimental information validating the existence of phase transition is presented.

1. Introduction

Although Ni crystals have been studied thoroughly, there are some inconsistent experimental data and commonly accepted theoretical concepts. Below, we discuss these inconsistencies and suggest a method of their elimination. The matter reduces to the following. At the Curie point $T_c$ =631 K, a nickel crystal is transformed into the ferromagnetic state. However, the constants of magnetic anisotropy have nonzero values only below *T*≈570 K [1]. In the range of 570 < *T* < 631 K, there is not even one vanishing constant of magnetic anisotropy and, hence, there are neither hard nor easy magnetization axes. In other words, within this temperature range, an isotropic (absolutely magnetically soft) ferromagnetic phase should exist. It is shown that this isotropic magnetic phase can be logically described within the Landau thermodynamic theory of phase transition. We consider the magnetic and structural transformations in nickel, following [2].

2. Choice of symmetry group of paramagnetic phase

In the paramagnetic phase, the nickel structure relates to the structural type of copper, *A*1, described by the sp. gr. $O_h^5$ [3]. The Ni atoms occupy the 1(a) positions (in Kovalev's notation [4]); these are cube vertices and face centers. The $O_h^5$ group is symmorphic, i.e., no screw axes and glide planes.

Thus, the space group of the paramagnetic phase can be chosen unambiguously. On the contrary, the magnetic symmetry group is chosen somewhat arbitrarily. It can be either the Shubnikov paramagnetic group $O_h^5 1'$ or the exchange paramagnetic group $O_h^5 \times O(3)$ [5], where 1' is the operation of spin inversion and O(3) is the three-dimensional rotation group in the spin space. In the first case, the existence of the isotropic phase cannot be described theoretically. Therefore, following Borlakov [6], we chose the sp. gr. $O_h$ x 0(3) possessing magnetic subgroups which coincide with the sp. gr. $O_h^5$.

3. Changes in magnetic symmetry in transition to the isotropic magnetic phase

In the Landau thermodynamic theory, the transition to a ferromagnetic phase is described by the magnetization vector $\vec{M}$. This order parameter behaves as a scalar with respect to the space transformations and is transformed as a vector in the spin space. Therefore, the order parameter $\vec{M}$ is transformed by the irreducible representation $A_{1g} \times V$ of the exchange paramagnetic group, where $A_{1g}$ is the unit irreducible representation of the group $O_h^5$ and $V$ is the vector irreducible representation of the group O(3). It should be indicated that the phase transitions occurring without the change of the translation symmetry can be described using, instead of the irreducible representation of the space group, the irreducible representation of the corresponding point group [7]. Since the symmetry of the crystal lattice and the ferromagnetic structure with respect to rotation about the direction of the magnetization vector $\vec{M}$ is preserved, the symmetry group of the isotropic phase is $O_h^5 \times O(1)$, where O(1) is the one-parametric group of rotations about the vector $\vec{M}$. Now, consider the transition from the isotropic to the anisotropic phase.

**4. Symmetry groups of anisotropic phases**

To determine the symmetry groups of anisotropic phases within the Landau theory, one has to know the irreducible representation of the group $O_h^5$, according to which the order parameter describing the phase transition is transformed. The order parameter, which describes the transition at the Curie point is transformed according to the irreducible representation $A_{1g} \times V$ of the magnetic group $O_h^5 \times O(3)$. The irreducible representation of the group $O_h^5$ describes the transition to the anisotropic phase and should be related to $A_{1g} \times V$. There is a method of successive derivation of a subgroup from the irreducible representation of a certain group which extends the restrictions of the initial irreducible representation to the subgroup [5]. The restriction of irreducible representation $A_{1g} \times V$ calculated by formulas from [5] is equal to the irreducible representation $F_{1g}$ of the group $O_h$. This agrees with [7], where the transition to the ferromagnetic phase in crystals of the class $O_h$ is induced by the irreducible representation $F_{1g}$ of the group $O_h$. The only difference in our case consists in the fact that this irreducible representation induces the transition from the isotropic to the anisotropic phase.

Denote the constraint imposed onto the irreducible representation of $A_{1g} \times V$ of the group $O_h$ x O(3) by the same symbol but enclosed into brackets, $\lfloor A_{1g} \times V \rfloor$. Then, the result can be written in the form $\lfloor A_{1g} \times V \rfloor = F_{1g}$. This irreducible representation of $O_h$ determines the lowering of the symmetry in the transition to the anisotropic ferromagnetic phase and is called critical [8].

Let the symmetry of a certain anisotropic phase be described by the point group $A_{1g} \times V$ and T(g) be the matrices of the irreducible representation corresponding to the elements $g \in G_D$. Since the structure of the anisotropic phase is invariant with respect to the group $G_D$, the following algebraic equation is valid for every element $g \in G_D$:

$$T(g)\vec{c} = \vec{c} \qquad (1)$$

Equation (1) signifies that the vector $\vec{c}$ is a stationary vector (S-vector) of the matrices of the irreducible representation. In other words, this vector is not changed by the matrices of the irreducible representation corresponding to the elements of the subgroup $G_D$. For each subgroup of the group $O_h$ there exists a specific S-vector; i.e., there is one-to-one correspondence between S-vectors of the irreducible representation and the subgroups of the initial group. The results of the calculations by Eq. (1) for various space groups are tabulated; thus, the necessary information can be taken, e.g., from Table 1. Obviously, the results of symmetry calculations for the star of the wave vector $\vec{k} = 0$ are the same for all the space groups of a crystallographic class.

**Table 1.** Low-symmetric phases induced by irreducible representation $F_{1g}$ of the group $O_h$.

| $\vec{c}$ | $ccc$ | $0cc$ | $c00$ | $c_1 c_2 c_3$ |
|---|---|---|---|---|
| $G_D$ | $C_{3i}(C_3)$ | $C_{2h}(C_2)$ | $C_{4h}(C_4)$ | $C_i(C_1)$ |
| EA | [111] | [011] | [100] | [lmn] |

The notation of magnetic point groups in the Table 1 was taken from [9]; i.e., the symbol of a magnetic group consists of two symbols of ordinary point groups and has the meaning which can be understood from the following example. The notation of the magnetic group of the trigonal phase is $C_{3i}(C_3)$, where $C_{3i}$ is the point group of the corresponding phase and $C_3$ is the subgroup of index 2 for the group $C_{3i}$. The black-white magnetic classes corresponding to the magnetically ordered states of a crystal should be designed just in this way [10]. The last row of table gives the symbols for easy-magnetization axes.

**5. Accompanying deformations**

The critical irreducible representation $F_{1g}$ determines the possible lowering of the symmetry reduction at the point of a phase transition. However, in addition to symmetry lowering and the appearance of hard- and easy-magnetization axes, some other physical phenomena consistent with the symmetry of a new phase [8] can occur in each anisotropic phase. Among them, there are accompanying spontaneous deformations. Both critical and noncritical irreducible representations form the so-called complete condensate. Performing computations by the scheme [8], one can obtain the complete condensate of the critical irreducible representation $F_{1g}$ of the group $O_h$ (Table 2).

Discuss the physical meaning of individual degrees of freedom involved in the complete condensate. The unit irreducible representation $A_{1g}$ describes the isotropic expansion or compression, accompanying the relativistic phase transition. The appropriate secondary order parameter is equal to the sum of the diagonal elements of deformation tensor [11]

$$a = u_{xx} + u_{yy} + u_{zz} \ . \tag{2}$$

The two-dimensional irreducible representation $E_g$ describes the compression-extension deformations accompanying the relativistic phase transition. The components of the order parameter are expressed in terms of the deformation-tensor components as follows:

$$a_1 = \frac{1}{\sqrt{6}}(2u_{zz} - u_{xx} - u_{yy}); \qquad a_2 = \frac{1}{\sqrt{2}}(u_{yy} - u_{xx}). \qquad (4)$$

The three-dimensional irreducible representation $F_{2g}$ with the components of the secondary-order parameter

$$a_1 = u_{yz}; \qquad a_2 = u_{zx}; \qquad a_3 = u_{xy} \qquad (5)$$

describes the shear deformations accompanying the transition to the anisotropic phase. The correlation between the data in Table 2 and the data obtained by formulas (2)-(4) is obvious.

Now we can proceed to the discussion of the results obtained for nickel crystals.

**Table 2.** Complete condensate of stationary vectors of the critical irreducible representation $F_{1g}$ of the group $O_h$

| $\vec{c}$ | $A_{1g}$ | $A_{2g}$ | $E_g$ | $F_{2g}$ |
|---|---|---|---|---|
| $ccc$ | $a$ | $a$ | - | $aaa$ |
| $0cc$ | $a$ | - | $a,0$ | $aab$ |
| $c00$ | $a$ | - | $a,-\sqrt{3}a$ | - |
| $c_1c_2c_3$ | $a$ | $a$ | $a,b$ | $abc$ |

**6. Discussion**

At room temperature, nickel is a ferromagnetic with an easy-magnetization [111] axis. However, nickel cannot be directly transformed from the isotropic to the trigonal phase. The first magnetic-anisotropy constant $K_1$ changes the sign at a temperature of T= 380 K [12]. Within the range of 380 < $T$ < 570 K, the constant $K_1 > 0$, whereas at $T$ < 380 K, the constant $K_1 < 0$. It is known [10] that the change of the sign of $K_1$ leads to the spin-flip phase transition with the change of the direction of the easy-magnetization axis.

Thus, during cooling of a nickel crystal from the Curie point, $T_c$ = 631 K, the following processes develop. At $T_c$, nickel is transformed to the isotropic ferromagnetic phase; i.e., the symmetry of crystal lattice remains the same, whereas the magnetization vector $\vec{M}$ can have an arbitrarily direction with respect to the crystallographic axes. At $T_{ls} \approx 570$ K, the crystal undergoes the transition to the anisotropic phase with the easy-magnetization axis along [100], because both magnetic-anisotropy constants are positive. The symmetry of crystal lattice is described by the sp. gr. $C_{4h}^5$. This phase transformation is a specific spin-flip transition from an arbitrarily directed easy-magnetization axis to the fourfold [100] axis. With the further lowering of the temperature down to 380 K, the constant $K_1$ becomes negative and the transition from tetragonal to rhombohedral phase with easy magnetization [111] axis occurs. The crystal-lattice symmetry of this phase is $C_3$, and is preserved to room temperature and even lower.

Note that the sequence of phase transitions obtained

$$O_h^5 \xrightarrow{631K} O_h^5 \xrightarrow{570K(470K?)} C_{4h}^5 \xrightarrow{380K} C_{3i}^2 \qquad (6)$$

in this article is inconsistent with the known data. According to [3, 13], the tetragonal nickel phase is similar to the tetragonal indium phase (A6) described by the symmetry group $D_{4h}^{17}$. The rhombohedral phase also has a higher symmetry $D_{3d}^5$ typical of the crystal lattice of mercury (A10). Thus, we face a contradiction, which, at first glance, is difficult to eliminate. However, our scheme provides the interpretation of all the symmetry groups determined experimentally and following from the group-theoretical considerations. Indeed, the symmetry of possible low-symmetry phases is determined by the critical irreducible representation $F_{1g}$ of the group $O_h^5$. However, all possible changes compatible with the symmetry of a newly formed phase can occur in a nickel crystal [8]. These additional changes are associated with accompanying effects, in particular, with deformation in the crystal.

The compression-extension deformation induced by the secondary irreducible representation $E_g$ gives rise to displacements characterized by the symmetry group $D_{4h}^{17}$. These displacements are small in comparison with the critical ones occurring in the vicinity of the temperature $T_{ls} \sim 570$ K of the relativistic phase transition. However, with an increase of the distance of the transition point, the noncritical displacements associated with compression-extension deformations start exceeding the critical values. Then, the X-ray diffraction analysis yields the symmetry group $D_{4h}^{17}$ corresponding to the secondary order parameter. This does not contradict the chain of transitions suggested in this study. Analogously, the trigonal displacements induced by the secondary irreducible representation $F_{2g}$ below 380 K exceed the critical displacements, so that, instead of the true symmetry $C_{3i}^2$, the X-ray diffraction analysis yields a higher symmetry $D_{3d}^5$. Note that the true temperature of the relativistic transition from the isotropic to the anisotropic phase seems to be lower than 570 K corresponding to the vanishing of the magnetic-anisotropy constants. The point is that the magnetic-anisotropy constants are measured in strong external magnetic fields, which fact, as is well known [14], shifts the transition point to higher temperatures. It is more probable that the point of relativistic phase transition has the temperature $T=T_{ls}=470$ K at which the pronounced λ-peak is observed on the curve of initial magnetic permeability of nickel [15]. Within our model, the formation of this peak is quite clear. Indeed, according to Kersten, the initial permeability in the anisotrpic phase is inversely proportional to the square root of the magnetic-anisotropy constant [15]: $\chi_a = cMK_1^{-\frac{1}{2}}$. Magnetization *M* is determined mainly by the exchange interaction and does not exhibit any anomalous changes near the relativistic transition. It should be noted that a small kink is formed on the dependence *M(T)*, which is explained by relativistic contribution to magnetization below the point $T_{ls}$ [16]. However, the magnetic-anisotropy constant tends to zero with the temperature rises up to $T_{ls} \sim 470$ K [16]. Thus, the initial permeability shows an anomalous increase. Note that we observed a feebly marked λ - peak at a temperature about 470 K by the method of differential thermal analysis [17]. The temperature dependence of the lattice parameter, has a kink at the above temperature [16], indicating a second-order phase transition at this temperature.

Thus, the set of experimental data indicate that nickel crystals are likely to have an isotropic phase within a temperature range of 470 < *T* < 631 K.

**7, Conclusion**

The changes in the magnetic symmetry, the temperature dependence of magnetic-anisotropy constants, magnetization, initial permeability, and some other physical characteristics of nickel crystals can be consistently interpreted on the basis of the Landau theory of phase transitions and the use of a postulate stating the existence of the isotropic phase from the temperature below the Curie point down to $T_{ls}\sim$ 470 K. At this temperature, the transition to the anisotropic phase takes place which is caused by the cooperative effect of relativistic interactions. At the Curie point, only the magnetic symmetry is lowered, while the symmetry of crystal lattice is preserved so that the Landau-Lifshitz criterion holds for the symmetry group of the isotropic phase [18].

Note that the above effects are more pronounced in crystals containing magnetic ions with triply degenerate orbital states [19]. In this instance, the relativistic interactions reduce to spin-orbit interactions. The cooperative effect of the relativistic interactions becomes much more pronounced than in pure 3d metals. However, it is usually interpreted as the cooperative Jahn-Teller effect and is considered independently of magnetic phenomena in crystals [20].

_________________________________

[*] Electronic address: borlakov@mail.ru